\documentclass[pre,twocolumn,amsmath,amssymb,superscriptaddress]{revtex4-2}

\usepackage{amsmath}
\usepackage{hyperref}
\usepackage{graphicx}
\usepackage{amsfonts}
\usepackage{amsthm}
\usepackage{cases}
\usepackage{bm}

\usepackage{color}
\definecolor{Blue}{rgb}{0.00, 0.00, 1.00}
\definecolor{Red}{rgb}{1.00, 0.00, 0.00}
\definecolor{Green}{rgb}{0.00, 0.70, 0.00}

\newcommand{\ve}[1][K]{\mathbf{#1}}
\hypersetup{
    colorlinks=true,       
    linkcolor=blue,          
    citecolor=blue,        
    filecolor=magenta,      
    urlcolor=cyan           
}

\begin{document}

\title{Self-phoretic oscillatory motion in a harmonic trap}
\author{Arthur  \surname{Alexandre}}
\affiliation{Laboratory of Computational Biology and Theoretical Biophysics, Institute of
 Bioengineering, School of Life Sciences, \'Ecole Polytechnique F\'ed\'erale de Lausanne}
\affiliation{Univ. Bordeaux, CNRS, LOMA, UMR 5798, F-33400, Talence, France.}
\author{Leah  \surname{Anderson}}
\affiliation{Univ. Bordeaux, CNRS, LOMA, UMR 5798, F-33400, Talence, France.}
\author{Thomas  \surname{Collin-Dufresne}}
\affiliation{Univ. Bordeaux, CNRS, LOMA, UMR 5798, F-33400, Talence, France.}
\author{Thomas  \surname{Gu\'erin}}
\affiliation{Univ. Bordeaux, CNRS, LOMA, UMR 5798, F-33400, Talence, France.}
\author{David S. \surname{Dean}}
\affiliation{Univ. Bordeaux, CNRS, LOMA, UMR 5798, F-33400, Talence, France.}
\affiliation{Team MONC, INRIA Bordeaux Sud Ouest, CNRS UMR 5251, Bordeaux INP, Univ. Bordeaux, F-33400, Talence, France.}
\date{\today}

\begin{abstract}
We consider the motion of a harmonically trapped overdamped  particle, which is submitted to a self-phoretic force, that is proportional to the gradient of a diffusive field for which the particle itself is the source. In agreement with existing results for free particles or particles in a bounded domain, we find that the system exhibits a transition between an immobile phase, where the particle stays at the center of the trap, and an oscillatory state. We perform an exact analysis giving access to the   bifurcation threshold, as well as the frequency of oscillations and their amplitude near the threshold. Our analysis also characterizes the shape of two-dimensional oscillations, that take place along a circle or a straight line. Our results are confirmed by numerical simulations. 
\end{abstract}

\maketitle

\section{Introduction}

In a range of physical situations, a single particle can produce a field which influences its motion. Perhaps the first example  in the physics literature is the Abraham-Lorentz force, which is due  to the interaction of the electromagnetic field generated by a charged particle, which itself acts on the charged particle. The results of Abraham and Lorentz were then extended to a fully relativistic setting by Dirac in the form of the Abraham-Lorentz-Dirac force \cite{dir38}.  In the aforementioned classical analysis, the mathematical challenge lies in solving simultaneously the Maxwell equations in the presence of a source given by the charged particle, in conjunction with the relativistic equation of motion of the particle subjected to its own electric field. 

A very similar problem is ubiquitous in the general field of active matter. Here, a particle is subjected to a phoretic force generated by gradients of a  field for which the particle is itself the source. For example, bacteria, or more generally microorganisms, tend to move along gradients of chemicals  which are secreted by the microorganisms themselves \cite{taxis,kel70}. Clearly, under certain circumstances, the chemical emitted by a sole bacterium leads to a chemical gradient which depends on the past trajectory of the microorganism~\cite{tsori2004self,gri05,grima2006phase,sen09,sen09b,new04,lie19}. Another example is the motion of a synthetic ``swimmer'' which experiences forces in the direction of the  gradient  of a chemical \cite{chemo}, which again may be emitted by the swimmer itself, so that the particle interacts with its own trail, examples include  camphor boats~\cite{koy16,koy15,koy19,miy17,nag04,nis15} and more generally self-propelled swimming droplets~\cite{jin2017chemotaxis,hokmabad2021emergence,maass2016swimming,michelin2013spontaneous}. 
 
In this context, a simplified generic model describing this self-phoretic motion is \cite{gri05,new04,lie19,sen09,sen09b,kra16}
\begin{equation}
\frac{d{\bf X}_t}{ dt}= -\lambda\nabla\phi({\bf X}_t)+\sqrt{2D_p}{\boldsymbol\eta}(t),\label{eqp}
\end{equation}
where $\ve[X]_t$ denotes the position of the particle at time $t$, $\lambda$ is a phoretic coefficient (or a measure of the activity of the system), and ${\boldsymbol\eta}(t)$ denotes normalized Gaussian white noise satisfying $\langle\eta_i(t)\eta_j(t')\rangle=2\delta_{ij}\delta(t-t')$ with $i,j$ the spatial coordinates, and $D_p$ is the diffusion constant when $\lambda=0$. The field $\phi$ itself obeys 
\begin{equation}
 \partial_t \phi({\bf x},t) = D\nabla^2\phi({\bf x},t) -\mu \phi({\bf x},t)+ \delta_\sigma({\bf x}-{\bf X}_t),\label{eqf}
\end{equation}
which is simply the diffusion equation with a diffusion constant $D$, with a degradation or decay rate $\mu$, and a source term centered at the particle position which we will take to be 
\begin{equation}
\delta_\sigma(\ve[x])= e^{-\ve[x]^2/(4\sigma^2)}/(4 \pi \sigma^2)^{\frac{d}{2}}, \label{GaussianSource}
\end{equation}
where $d$ denotes the spatial dimension and $\sigma$ is an effective particle size. 
The field $\phi$ is usually taken to be a  concentration field, however in principle it could  also be a temperature field which induces motion via thermophoresis \cite{thermo}, in which case $\lambda$ would be the  Soret coefficient of the particle that would be subject to a radiation source  from which it absorbs energy and heats up (while the absorption by the background solvent is negligible). 
One can also see  Eq.~(\ref{eqp}) as a linear response relation, expressing the fact that the existence of a non-zero gradient in the field $\phi$  means that the system is out of thermodynamic equilibrium, and must then   generate particle motion to drive the system towards thermodynamic equilibrium.  Note that in Eq.~(\ref{eqp}), if $\phi$ is a temperature field, the dependence of the diffusion constant (via the viscosity) on temperature is neglected, in contrast with the ``hot Brownian motion'' model where such temperature-dependent viscosity  is taken to be the dominant effect \cite{hotbm}.

The model given by Eqs.~(\ref{eqp}) and (\ref{eqf})  has been extensively studied \cite{gri05,new04,lie19,sen09,sen09b,kra16}, sometimes with variations, such as including inertia, or considering a time delay in the source term (representing the measurement time of the microorganism) instead of a particle size~\cite{gri05}. In the absence of noise, this model shows a spontaneous symmetry breaking \cite{lie19,gri05} when $\lambda$ exceeds a positive threshold value. In this case the particle is so repelled by the trail it emits that it spontaneously sets into motion at a finite velocity. In the presence of noise, the particle can change direction and in two dimensions has a behavior very similar to that of an active Brownian particle (ABP) \cite{romanczuk2012active}, where the speed is more or less constant and the direction of motion diffuses. 

In this paper, we investigate the question of how such self-phoretic particles behave in confinement. Since free particles acquire a spontaneous velocity above a threshold, we may expect that  confined particles do not rest at an equilibrium position but rather escape this position and display oscillatory motion, as is the case in other contexts of molecular motor assemblies \cite{Juelicher1997PRL,Placais2009,Guerin2011b}, cilia beating~\cite{camalet1999self} or oscillations of cellular protrusions~\cite{sens2020stick}. Indeed, oscillatory motion has been experimentally observed in the context of camphor boats confined in one dimensional  (narrow slit) and two-dimensional (disk) domains \cite{hayashima2001camphor,koy16,koy15,koy19,miy17,nag04,nis15,bon19}. 
We also note that,  experiments where active particles are harmonically confined using acoustic  \cite{tak16}  or optical \cite{schmidt2021non} traps or parabolic surfaces~\cite{dau19} show regimes where particles tend to avoid the trap's center. However, descriptions of motion of active particles in confinement rely either on  the use of phenomenological models, such as run and tumble models, ABPs, etc \cite{tai09,pot12,wag17,das18,cha21,dha19,bas20,mal20,dau19}, or on a time derivative expansion valid only for a rapid relaxation of the field $\phi$ \cite{koy16,koy15,koy19}. The purpose of the present work is to present exact analytical results for harmonically confined self-phoretic motion, that do not rely on such rapid-relaxation assumptions or on effective models, in order to take into account the full non-Markovian feature of the dynamics of $\ve[X]_t$. 

The outline of this paper is as follows. First, we derive an effective equation of motion for $\ve[X]_t$ (Section \ref{SecEqMot}). Then, we carry out a linear stability analysis where we identify an oscillatory instability (Hopf bifurcation) and we identify exactly the phase diagram of the system for any spatial dimension (Section \ref{SecLinear}). Then, we calculate asymptotically exact formulas for both the amplitude and frequency of the weakly non-linear oscillations, slightly above the bifurcation threshold (Section \ref{SecNonLin}). Furthermore, our analysis also explains how higher harmonic terms are generated as the activity increases. The results described above are largely explained by an exact analytical  treatment close to the transition and in the case where the particle {\em size}  $\sigma$ is taken to zero particularly explicit results can be found. In two-dimensions we also describe the shape of the oscillations, which describe circles in the two-dimensional  plane instead of straight lines.  Finally, we present numerical simulations of the model in Section \ref{numerics} and present concluding remarks in Section \ref{SecConclusion}.

\section{Equation of motion}
\label{SecEqMot}
In this paper we analyze the motion described by equations  (\ref{eqp}) and (\ref{eqf}), but where we introduce an additional  harmonic trapping force:
\begin{align}
& \dot{\ve[X]_t}=-K\ \ve[X]_t - \lambda\nabla\phi+\sqrt{2D_p}{\boldsymbol\eta}(t), \\
&\partial_t\phi=D\nabla^2\phi -\mu \phi +\delta_\sigma(\ve[x]),
\end{align}
where $K$ represents the stiffness of a harmonic potential (renormalized by the friction coefficient, so that it has the dimension of an inverse of time). If one knows the trajectory $\ve[X]_t$ up to $t$, at all past times, then the field $\phi$ at a position $\ve[x]$  is a sum of the effects of all past sources, whose intensity decays at the rate $\mu$, so that
\begin{align}
\phi(\ve[x],t)=\int_0^\infty d\tau \int d\ve[y] \ \frac{e^{-\frac{(\ve[x]-\ve[y])^2}{4D\tau}-\mu\tau}}{(4\pi D\tau)^{d/2}} \delta_\sigma(\ve[y]-\ve[X]_{t-\tau}),
\end{align}
which is obtained using the Green's function of the diffusion equation. For the Gaussian source $\delta_\sigma$ given by (\ref{GaussianSource}), this leads to 
\begin{align}
\phi(\ve[x],t)=  \int_0^\infty d\tau  \frac{e^{-\frac{(\ve[x]-\ve[X]_{t-\tau})^2}{4(D\tau+\sigma^2)}-\mu\tau}}{[4\pi (D\tau+\sigma^2)]^{d/2}}.  
\end{align}
Using this expression to calculate $\nabla\phi$, we obtain the effective equation of motion for $\ve[X]$ as
\begin{align}
&\dot{\ve[X]}_t=-K\  \ve[X]_t + \sqrt{2D_p}{\boldsymbol\eta}(t) \nonumber\\
&+    \int_0^\infty d\tau (\ve[X]_t-\ve[X]_{t-\tau}) \frac{2 \lambda\ e^{-\frac{(\ve[X]_t-\ve[X]_{t-\tau})^2}{4(D\tau+\sigma^2)}-\mu\tau}}{ \pi^{d/2}[4 ( D\tau+\sigma^2)]^{d/2+1}}. \label{EqMotion}
\end{align}
This equation shows that $\ve[X]_t$ is a strongly non-Markovian process, since the future evolution depends on the whole past trajectory. We also see that, in the absence of noise, $\ve[X]_t=\ve[0]$ is a trivial solution of the problem, which may however be neither unique nor stable. Similar equations have been studied in Refs.~\cite{gri05,sen09}, and in the absence of confinement ($K=0$) a threshold of $\lambda$ was identified beyond which the particle acquires a finite velocity after a spontaneous symmetry breaking. This was found for vanishing particle sizes, with a time delay to regularize the integrals \cite{sen09} which is equivalent to our case with finite particle sizes.
In the next sections, we turn to the case with confinement ($K>0$) and perform an exact study of the stability of the trivial solution $\ve[X]_t=\ve[0]$ and find the location at which an oscillatory instability appears.

\section{Oscillatory instability: linear analysis}
\label{SecLinear}
\subsection{Location of the bifurcation}
Keeping only terms that are linear in $\ve[X]_t$ in Eq.~(\ref{EqMotion}) and going to Fourier space (with the convention  $\hat{f}(\omega)=\int_{-\infty}^\infty dt f(t) e^{-i\omega t}$), we obtain
\begin{align}
&i\omega  \hat{\ve[X]}(\omega)=-K \  \hat{\ve[X]} (\omega)+\lambda \mathcal{G}(\omega) \hat{\ve[X]}(\omega)  + \sqrt{2D_p}\hat{{\boldsymbol\eta}}(\omega),
\end{align}
where we have defined
\begin{align}
\mathcal{G}(\omega)= \int_0^\infty d\tau\frac{2(1-e^{-i\omega \tau})e^{-\mu\tau}}{ \pi^{d/2}[4(\sigma^2+D\tau)]^{d/2+1}}.
\end{align}
We may write this equation as 
\begin{align}
\hat{\ve[X]}(\omega)= {\mathcal{R}}(\omega) \sqrt{2D_p}\hat{{\boldsymbol\eta}}(\omega)=\frac{ \sqrt{2D_p}}{i\omega+K-\lambda\mathcal{G}(\omega)}  \hat{{\boldsymbol\eta}}(\omega), \label{RespFunction}
\end{align}
where  $\mathcal{R}$ is the linear response function. If the steady state solution is finite at all times, then causality implies (given the convention of Fourier transforms used here) that all the poles of $ {\mathcal{R}}(\omega)$ are in the upper complex plane. The immobile solution $\ve[X]=\ve[0]$  thus becomes unstable when the response function has a pole on the real axis. This means that there is a bifurcation when one of the poles of ${\mathcal{R}}$ crosses the real axis at some frequency $\omega_c$ (and another pole also crosses the real axis, at frequency $-\omega_c$). This occurs for a critical value of the coupling $\lambda=\lambda_c$, which can be identified from the equation
\begin{align}
i\ \omega_c   =- K  +\lambda_c \mathcal{G}(\omega_c) \label{EqBifurcationComplex},
\end{align}
which can be also written as
\begin{align}
&\lambda_c=\omega_c/\mathcal{G}_i(\omega_c),\label{B1}\\
&K=\omega_c   \mathcal{G}_r(\omega_c)/\mathcal{G}_i(\omega_c),\label{B2}
\end{align}
where $\mathcal{G}_r$ and  $\mathcal{G}_i$ are respectively the real and the imaginary parts of $\mathcal{G}$.  These equations define, in an implicit form, the location of the Hopf bifurcation. More precisely, this defines a curve in the $(K,\lambda)$ plane that is parametrized by $\omega_c$, so that the phase-diagram can be straightforwardly obtained.

\begin{figure}
    \includegraphics[width=\linewidth]{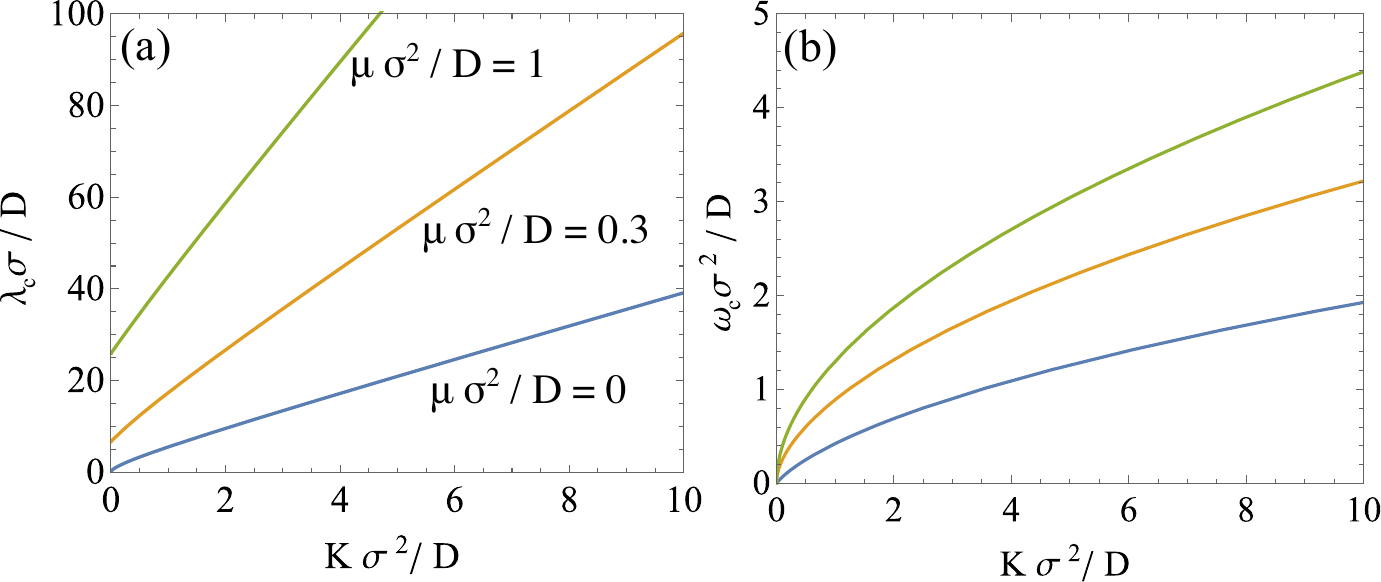}
    \caption{(a) Behavior of line of transitions between oscillating and non-oscillating phases in the plane $(K,\lambda_c)$ for $d=1$. Here the units are set so that $D=\sigma=1$ and one shows different critical curves correspond to different values of $\mu$. (b) Unstable frequency $\omega_c$ at the transition as a function of $K$ for the same parameters. }  \label{FigEffectMu}
\end{figure}

\begin{figure}
    \includegraphics[width=\linewidth]{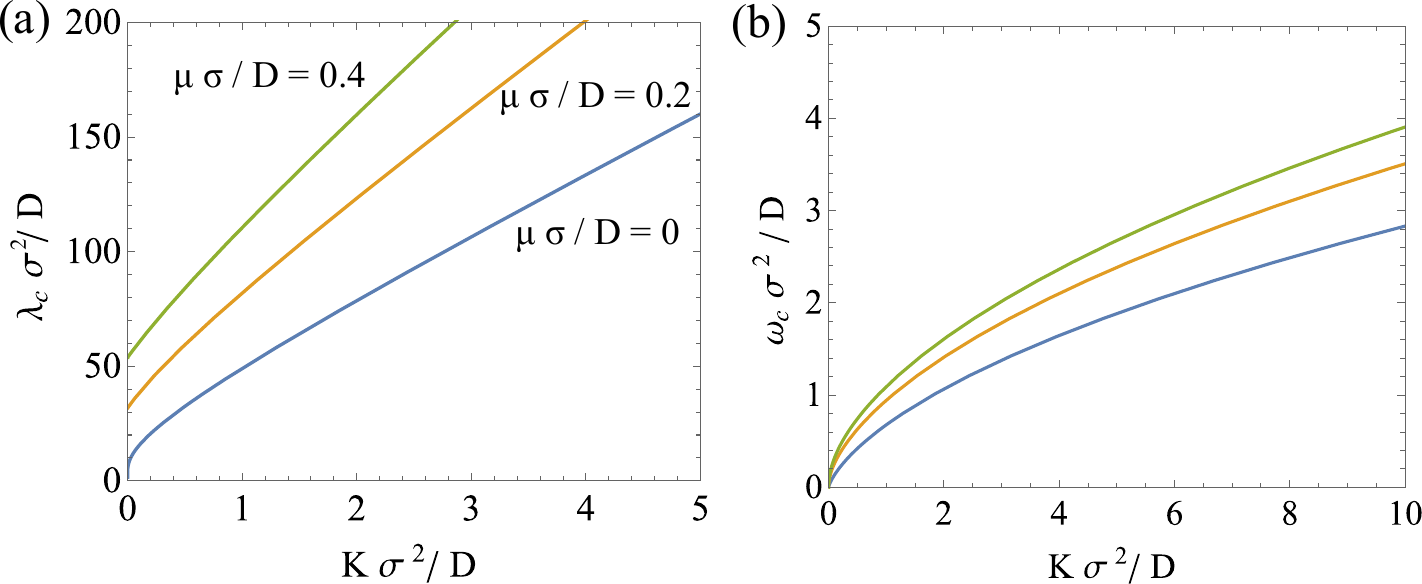}
    \caption{(a) Behavior of line of transitions between oscillating and non-oscillating phases in the plane $(K,\lambda_c)$ for $d=2$. Here the units are chosen so that $D=\sigma=1$ and one shows different critical curves correspond to different values of $\mu$. (b) Unstable frequency $\omega_c$ at the transition as a function of $K$ for the same parameters. }  \label{FigEffectMu2D}
\end{figure}

\begin{figure}
    \includegraphics[width=\linewidth]{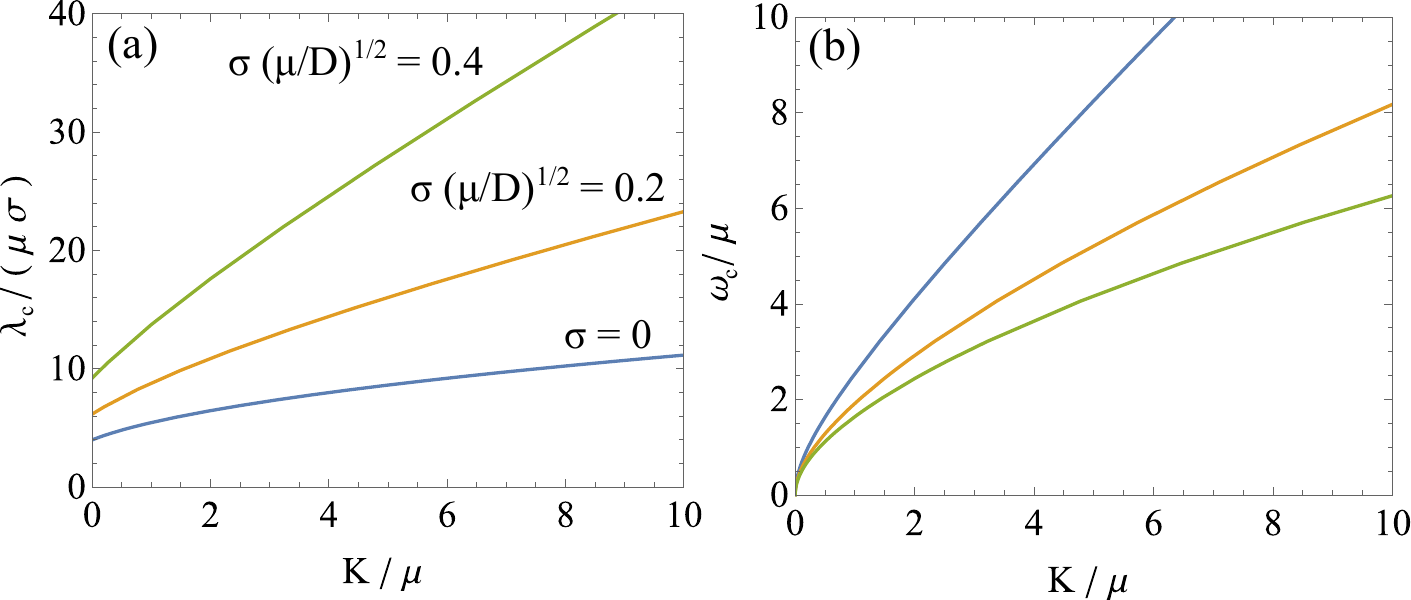}
    \caption{(a) Line of transitions between oscillating and immobile states in the plane $(K,\lambda_c)$ for $d=1$. Here the units are set so that $D=\mu=1$ and one shows different critical curves correspond to different values of $\sigma$. 
     (b) Unstable frequency $\omega_c$ at the transition as a function of $K$ for the same parameters.}  \label{FigEffectSigma}
\end{figure}

In Fig.~\ref{FigEffectMu}(a), we show how the phase boundary changes for fixed values of $D$ and $\sigma$ as the evaporation term $\mu$ is varied. As $\mu$ is decreased the memory effect becomes stronger and the threshold value  $\lambda_c$ is reduced, which can be physically understood since for large evaporation rate the field vanishes too quickly to obtain an instability.  Similar behaviors are obtained for $d=2$, see Fig.~\ref{FigEffectMu2D}. Next, in Fig.~\ref{FigEffectSigma}, we show how the phase boundary changes for fixed values of $D$ and $\mu$ as the effective particle size $\sigma$ is varied. As $\sigma$ is increased we expect that this suppresses gradients in the field $\phi$ and thus the transition is pushed to higher values of $\lambda_c$. In Fig.\ref{FigEffectSigma}(a), we see that this is indeed the case. However, the frequency of the unstable mode at the transition is actually increased as $\sigma$  decreases, as shown in Fig.~\ref{FigEffectSigma}(b). 

In all these curves, we see that $\omega_c$ vanishes when the (normalized) stiffness of the potential $K$ tends to zero, this can be understood analytically by noting that $\mathcal{G}_i(\omega)\sim \omega$, whereas $\mathcal{G}_r(\omega)\sim \omega^2$ for small $\omega$, so that 
\begin{align}
\omega_c \underset{K\to0}{\sim} \sqrt{K} .
\end{align}
For low values of the stiffness, the oscillations at the threshold are therefore very slow. For larger ones, the frequency increases, and so does the value of the threshold $\lambda_c$. This stabilization effect by the stiffness, due to suppression of memory effects, is well known in other contexts, such as stick slip oscillations \cite{batista1998bifurcations,carlson1996}, oscillations induced by molecular motors \cite{Guerin2011b} and oscillations of cellular protrusions \cite{sens2020stick}.

\subsection{Explicit formulas for a point-like particle in $d=1$.}
In one dimension, we can take the limit $\sigma\to 0$, this has the advantage of providing more explicit analytical results for the phase boundary and the frequency $\omega_c$ associated with the instability at the boundary. In this case we find that
\begin{align}
\mathcal{G}&(\omega_c) = \frac{1}{2{D^{\frac{3}{2}}}}\left[\sqrt{\mu+i\omega_c}-\sqrt{\mu}
\right], \nonumber\\
&= \frac{1}{2{D^{\frac{3}{2}}}}\left[(\mu^2+\omega_c^2)^{\frac{1}{4}}\left(\cos\frac{\theta}{2}+ i\sin\frac{\theta}{2}\right) -\sqrt{\mu}\right],
\end{align}
where
\begin{equation}
\theta = \tan^{-1}\left(\frac{\omega_c}{\mu}\right).
\end{equation}

Using the identities $\omega_c= (\mu^2+\omega_c^2)^{\frac{1}{2}}\sin\theta$ and $\sin\theta=2\sin\frac{\theta}{2}\cos\frac{\theta}{2}$, equating the imaginary parts of Eq.~(\ref{EqBifurcationComplex}) gives
\begin{equation}
(\mu^2+\omega_c^2)^{\frac{1}{4}}\cos\frac{\theta}{2}=\frac{\lambda_c}{4{D^{\frac{3}{2}}}} ,\label{imp}
\end{equation}
while equating the real parts leads to
\begin{equation}
K =\frac{\lambda_c}{2{D^{\frac{3}{2}}}}\left[(\mu^2+\omega_c^2)^{\frac{1}{4}}\cos\frac{\theta}{2} -\sqrt{\mu}\right]. \label{rp}
\end{equation}
Combining the above equations then gives an equation for $\lambda_c$ which can be solved, giving the phase boundary between oscillating  and non-oscillating phases as

\begin{equation}
\lambda_c =  2D^{\frac{3}{2}}\left(\sqrt{\mu} + \sqrt{2 K +\mu}\right).\label{lambdacPointLike}
\end{equation}
In the case where $K=0$, we recover the stability criterion given in Ref.~\cite{gri05} for a free particle in the limit where the time delay $\tau$ is taken to zero, which is equivalent to our model when $\sigma$ vanishes. 
In agreement with the previous numerical analysis for non-zero $\sigma$, we see that the value to $\lambda$ necessary to trigger the instability increases with $D$, $K$ and $\mu$.
Now, using the standard trigonometric formula 
\begin{equation}
\cos\frac{\theta}{2}=\sqrt{\frac{\cos\theta+1}{2}}=\sqrt{\frac{1}{2}}\sqrt{\frac{\mu}{\sqrt{\mu^2+\omega_c^2}}+1},
\end{equation}
along with  Eqs.~(\ref{imp}) and (\ref{lambdacPointLike}),  we find that the frequency at the onset of bifurcation is
\begin{equation}
\omega_c=\sqrt{K \left\{ K+ 2\left[\mu +\sqrt{\mu(\mu + 2K)}\right]\right\}}.\label{wcK}
\end{equation}
Interestingly, when $\mu=0$ the frequency associated with the instability is simply $\omega_c =K$. This analytical result is also in accordance with the numerical results for finite $\sigma$,  showing an increase of the unstable frequency on increasing $K$ and $\mu$.

We have thus determined the phase boundary between the active and passive phases and in particular, we have explicit analytical results for the case where $\sigma=0$ and $d=1$.

\section{Weakly non-linear oscillations}
\label{SecNonLin}
\subsection{Amplitude equation}

Now, we will analyze the behavior of the oscillating phase close to the transition. We will show that close to the transition, the oscillation frequency is given by $\omega_c$ determined in the previous section via the stability analysis. We also determine the continuous nature of the transition by computing the amplitude of the oscillations. Our analysis also shows how this first harmonic term then generates higher harmonics leading to deviations from a single mode of oscillation and the more complex behavior in the strongly oscillating regime.

Let us keep cubic terms in the equation (\ref{EqMotion}), where we again omit noise:
\begin{align}
&\dot{\ve[X]}_t\simeq-K\ \ve[X]_t +    \int_0^\infty d\tau  \frac{2\lambda (\ve[X]_t-\ve[X]_{t-\tau})  \ e^{-\mu\tau}}{ \pi^{d/2}[4 ( D\tau+\sigma^2)]^{d/2+1}}  \nonumber\\
&-    \int_0^\infty d\tau (\ve[X]_t-\ve[X]_{t-\tau})  \frac{2\lambda (\ve[X]_t-\ve[X]_{t-\tau})^2 e^{-\mu\tau}  }{ \pi^{d/2}[4 ( D\tau+\sigma^2)]^{d/2+2}}. \label{EqCubic}
\end{align}
Let us place ourselves in the vicinity of the bifurcation: $\lambda=\lambda_c(1+\varepsilon)$, with small $\varepsilon$. Following standard techniques in non-linear physics, we use the ansatz
\begin{align}
\ve[X]_t=\ve[A]_t\  e^{i \omega_c t}+c.c.
\end{align}
where $c.c.$ denotes the complex conjugate and $\ve[A]_t$ is a $d$-dimensional vector with complex components. Here, one assumes that $\ve[A]_t$ varies slowly compared to the timescale $\omega_c^{-1}$, this is justified since we will actually find that $\dot{\ve[A]}_t$ is of order $\varepsilon \ve[A]_t$.  
Now, since $\ve[A]_t$ varies slowly, we can use $\ve[A]_{t-\tau}\simeq\ve[A]_t-\tau\dot{\ve[A]}_t+...$ to evaluate the integrals in Eq.~(\ref{EqCubic}). The first term on the right hand side of this equation  then reads
\begin{align}
& \int_0^\infty d\tau  \frac{ 2 (\ve[X]_t-\ve[X]_{t-\tau})  \ e^{-\mu\tau}}{ \pi^{d/2}[4 ( D\tau+\sigma^2)]^{d/2+1}}  
  \nonumber\\
&\simeq 2\int_0^\infty d\tau    \frac{ e^{-\mu\tau}  [\ve[A]_t e^{i \omega_c t} - (\ve[A]_t-\dot{\ve[A]}_t\tau) e^{i\omega_c(t-\tau)}+c.c.] }{ \pi^{d/2}[4 ( D\tau+\sigma^2)]^{d/2+1}}  \nonumber\\
&\simeq e^{i \omega_c t}\left\{ \ve[A]_t \mathcal{G}(\omega_c)+\dot{\ve[A]}_t \mathcal{Q}(\omega_c)  \right\}+c.c.\label{Eval1}
\end{align}
where
\begin{align}
&\mathcal{Q}(\omega)=\int_0^\infty d\tau \frac{ 2 \tau e^{-\mu\tau-i\omega \tau}}{ \pi^{d/2}[4 ( D\tau+\sigma^2)]^{d/2+1}} =   i \partial_\omega\mathcal{G}(\omega).
\end{align}
We also have
\begin{align}
& \dot{\ve[X]}_t=(i\omega_c\ve[A]_t+\dot{\ve[A]}_t)e^{i\omega_ct}+c.c.\label{EvalDotX}
\end{align}
Last, the cubic term in Eq.~(\ref{EqCubic}) can be evaluated with $\ve[A]_{t-\tau}\simeq\ve[A]_t$, since higher orders will generate negligible terms. Omitting non-resonant terms (proportional to $e^{\pm3\omega_ct}$), we thus obtain
\begin{align}
& \int_0^\infty d\tau (\ve[X]_t-\ve[X]_{t-\tau})  \frac{ 2 (\ve[X]_t-\ve[X]_{t-\tau})^2 e^{-\mu\tau}  }{ \pi^{d/2}[4 ( D\tau+\sigma^2)]^{d/2+2}} \nonumber\\
& \simeq e^{i\omega_ct} \mathcal{W}(\omega_c)   \left\{ \ve[A]_t^*  (\ve[A]_t\cdot\ve[A]_t)   +2\ve[A]_t (\ve[A]_t\cdot\ve[A]_t^*)  \right\}+c.c.  
\label{EvalCub}
\end{align}
where we have defined
\begin{align}
\mathcal{W}(\omega_c)
=&\int_0^\infty d\tau    \frac{4   e^{-\mu\tau}(1-e^{-i\omega_c \tau}) [1-\cos(\omega_c\tau)]  }{ \pi^{d/2}[4 ( D\tau+\sigma^2)]^{d/2+2}}.
\end{align}
Collecting the results in Eqs.~(\ref{Eval1}), (\ref{EvalDotX}) and (\ref{EvalCub}), and using the equation (\ref{EqBifurcationComplex}) defining $\lambda_c$, Eq.~(\ref{EqCubic}) becomes
\begin{align}
\dot{\ve[A]}_t \simeq & \lambda_c \Big\{\varepsilon \ve[A]_t \mathcal{G}_c +\nonumber\\
&  \dot{\ve[A]}_t  \mathcal{Q}_c   - \mathcal{W}_c  [\ve[A]_t^*  (\ve[A]_t\cdot\ve[A]_t)   +2\ve[A]_t (\ve[A]_t\cdot\ve[A]_t^*)]  \Big\}, 
\end{align}
where, for conciseness we have written $\mathcal{G}_c=\mathcal{G}(\omega_c)$, $\mathcal{Q}_c=\mathcal{Q}(\omega_c)$, $\mathcal{W}_c=\mathcal{W}(\omega_c)$. Collecting the terms proportional to $\dot{\ve[A]}_t$ leads to
\begin{align}
&\dot{\ve[A]}_t = \lambda_c\frac{ \varepsilon \mathcal{G}_c \ve[A]_t  -  \mathcal{W}_c  [\ve[A]_t^*  (\ve[A]_t\cdot\ve[A]_t)   +2\ve[A]_t (\ve[A]_t\cdot\ve[A]_t^*)]   }{1-\lambda_c \mathcal{Q}_c} \label{EqAt},
\end{align}
which is the amplitude equation for our system.

\begin{figure}
    \centering
    \includegraphics[width=0.9\linewidth]{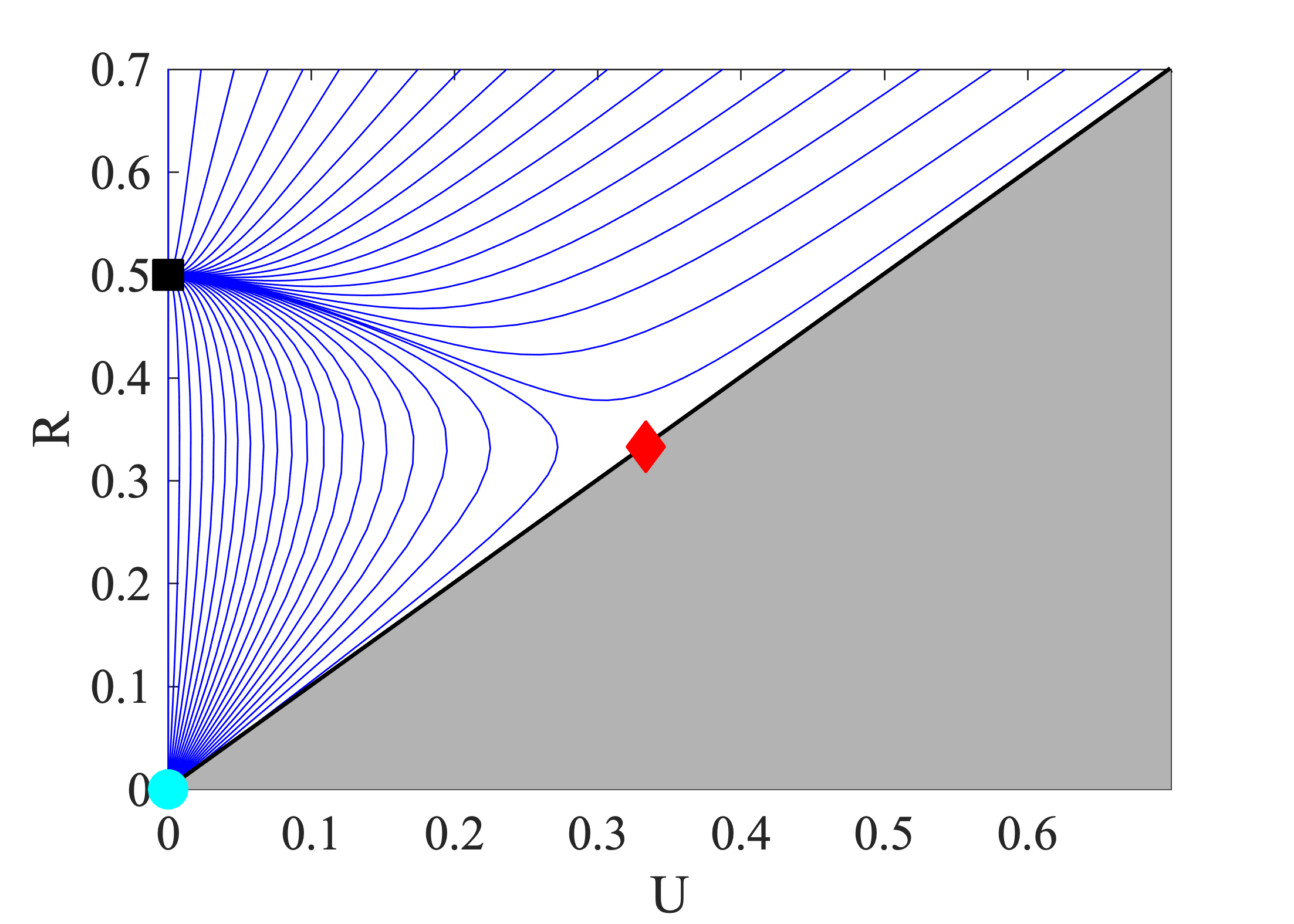}
    \caption{Flow lines for the dynamical system defined in Eqs.~(\ref{EqU}) and  (\ref{EqR}) above the bifurcation threshold for $\varepsilon\alpha=\beta=1$, which can be used without loss of generality by proper rescaling of $U$ and $R$. We show the three fixed points: the non-oscillating state (cyan disk), oscillations in a straight line (red diamond) and circular oscillations (black square). The gray area, with $R<U$, is physically forbidden.}  \label{TrajEnvelope}
\end{figure}

The variable $\ve[A]_t$ in this equation is a complex valued vector, and it is convenient to derive an equation for two positive scalar quantities  $R_t$ and $U_t$ defined by 
\begin{align}
&R_t=\ve[A]_t\cdot\ve[A]_t^*, 
&U_t=\sqrt{(\ve[A]_t\cdot\ve[A]_t)(\ve[A]_t^*\cdot\ve[A]_t^*)}. 
\end{align}
Using the amplitude equation for $\ve[A]_t$ (and its complex conjugate),  we obtain after some algebraic manipulations the equations describing the evolution of $U_t$ and $R_t$ (see Appendix \ref{EqUAndP}):
\begin{align}
&\dot{U}_t=U_t[\varepsilon\alpha     - 3 \beta    R_t  ], \label{EqU} \\
&\dot{R}_t =  \varepsilon \alpha  R_t - \beta  [U_t^2   +2R_t^2]  , \label{EqR}
\end{align}
where $\alpha$ and $\beta$ are real coefficients defined by
\begin{align}
&\alpha= \frac{\lambda_c\mathcal{G}_c}{1-\lambda_c \mathcal{Q}_c}   +c.c. , 
&\beta=\frac{\lambda_c  \mathcal{W}_c}{1-\lambda_c \mathcal{Q}_c}  +c.c., \label{DefAlphabeta}
\end{align}
which turn out to be both positive. 

The flow lines of the dynamical system formed by Eqs.~(\ref{EqU}) and (\ref{EqR}) are represented on Fig.~\ref{TrajEnvelope}. One identifies three fixed points. The first one is $(U=0,R=0)$ and corresponds to a non-oscillating state. The second fixed point is 
\begin{align}
U=R=\frac{\varepsilon\alpha}{3\beta}. \label{FP1}
\end{align}
This fixed point corresponds to oscillations that take place along a straight line in the $d-$dimensional space. The third fixed point exists only if the spatial dimension is $d>1$ and is   
\begin{align}
&U=0,&R=\frac{\varepsilon\alpha}{2\beta}, \label{FP2}
\end{align}
it corresponds to circular trajectories for two-dimensional systems, see Appendix
\ref{EqUAndP}. Looking at the flow lines on fig.~\ref{TrajEnvelope}, one realizes that for $\varepsilon>0$ the fixed point corresponding to circular oscillations is the only stable one, and it is straightforward to check it analytically. Interestingly, some lines approach very close to the fixed point corresponding to straight trajectories, so that one may expect that the system may exhibit temporary trajectories in straight line before converging to the final state with circular oscillations.

\subsection{Frequency shift and anharmonicity of the oscillations}
The frequency of the dominant mode near the transition corresponds to the frequency of the instability leading to the transition, with  possibly a small shift when the bifurcation threshold is crossed. If there is a frequency shift $\Delta \omega$, it means that in the steady state, $\ve[A]_t\sim e^{i\Delta\omega t}$ and therefore $\dot{\ve[A]}_t=i\Delta\omega\ve[A]_t$. Inserting this into (\ref{EqAt}), which we multiply by $\ve[A]_t^*$, leads to
\begin{align}
& \Delta\omega   =\frac{\lambda_c}{i R} \left[ \frac{ \varepsilon \mathcal{G}_c R  -   \mathcal{W}_c  (U^2   +2  R^2)   }{1-\lambda_c \mathcal{Q}_c}\right].
\end{align}
This formula provides the frequency shift of the oscillations  as one leaves the critical line  in terms of the single parameter $\omega_c$. 

Up to now we have considered only perfectly harmonic oscillations, whose amplitude (along each spatial coordinate) varies as $e^{\pm i\omega_c t}$. However, cubic non-linearities also generate terms of the form $e^{\pm 3 i\omega_c t}$. To quantify this effect, we consider solutions of the form
\begin{align}
\ve[X]_t=\ve[A]_t\ e^{i \omega_c t} + \ve[A]_t^{(3)}\ e^{i3 \omega_c t}+c.c.,
\end{align}
which we insert into (\ref{EqCubic}) to select terms of frequency $3\omega_c$, thereby obtaining
\begin{align}
\ve[A]_t^{(3)} = - {\mathcal{R}}(3\omega)\lambda_c \mathcal{S}_c \ve[A]_t  (\ve[A]_t\cdot\ve[A]_t), \label{AnHarm}
\end{align}
where the response function $\mathcal{R}$ was defined in Eq.~(\ref{RespFunction}), and 
\begin{align}
\mathcal{S}_c= \int_0^\infty d\tau    \frac{  2 e^{-\mu\tau}(1-e^{i\omega_c \tau})^3}{ \pi^{d/2}[4 ( D\tau+\sigma^2)]^{d/2+2}}.
\end{align}
Note that $\mathcal{R}(\omega)$ has a pole at $\omega=\omega_c$ when $\lambda=\lambda_c$, but $\mathcal{R}(3\omega_c)$ is finite. Hence, (\ref{AnHarm}) predicts the presence of anharmonic terms with the triple frequency, of the order $A^{(3)}\sim A^3\simeq \varepsilon^{3/2}$. 
Noting that $A\propto \sqrt{\varepsilon}$, this means that those anharmonic terms are one order smaller than the leading order harmonic contributions in the solution for $\ve[X]_t$. Another interesting remark is that for the fixed point corresponding to $d$-dimensional oscillations, we have $U=0$ and therefore $\ve[A]\cdot\ve[A]=0$, so that Eq.~(\ref{AnHarm}) indicates that the anharmonicity \textit{vanishes} for such circular oscillations. In turn, in $d=1$, where oscillations occur on a line, the anharmonicity does not vanish and one may therefore expect oscillations that differ from a sinusoidal shape when one leaves the threshold.    

\begin{figure}
    \centering
    \includegraphics[width=\linewidth]{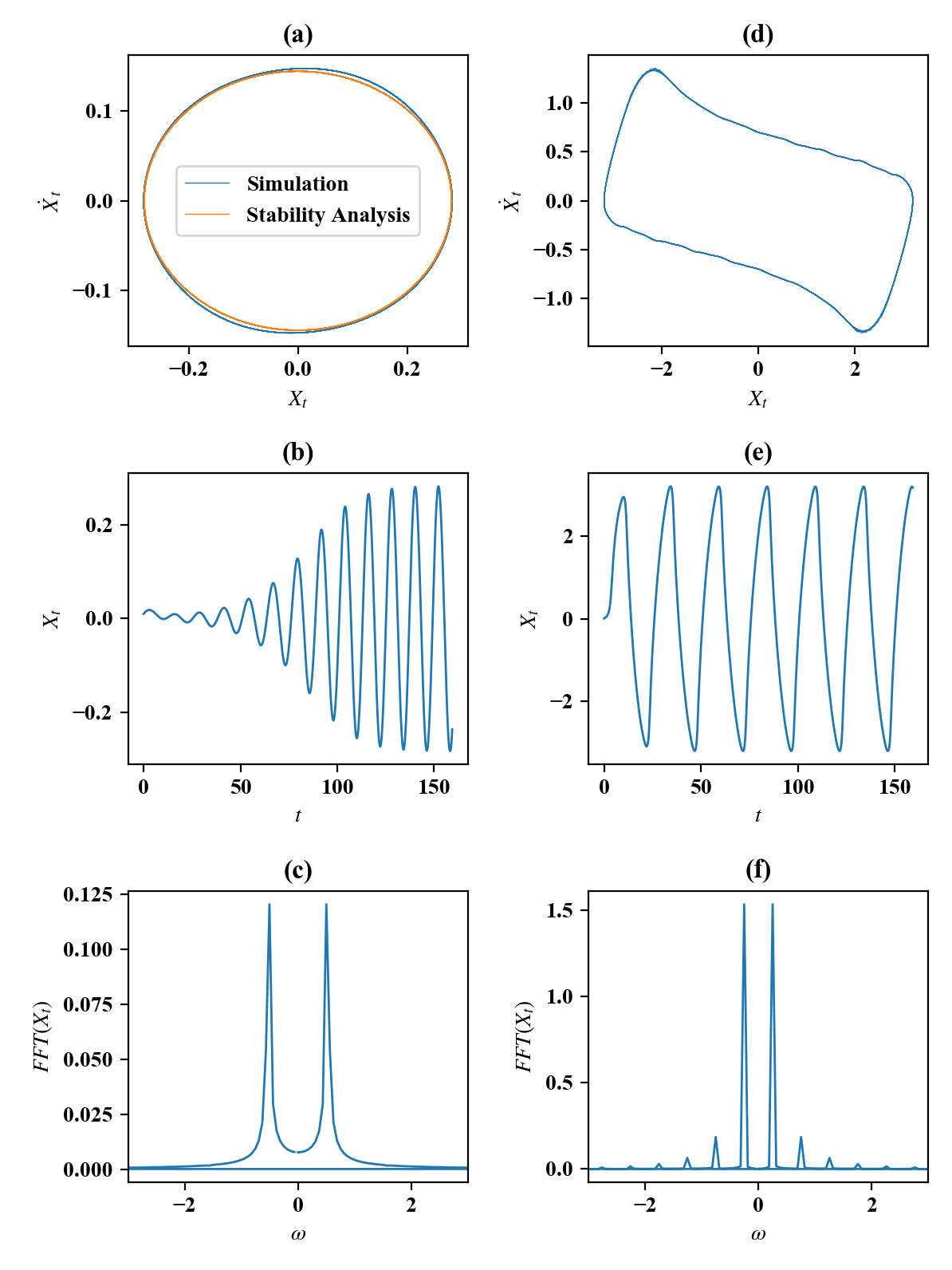}
        \caption{Numerical oscillations for the one dimensional problem obtained with the method of Section \ref{1DSimu}. In (a),(b),(c), we used $\mu=0.1$, $D=0.1$, $\sigma=0.3$, $K=0.6$, $L=10$ and $\lambda=0.25$. The value of the threshold for these parameters is $\lambda_c\simeq 0.236$. (a): trajectories in the $(X_t,\dot{X}_t)$ plane, the blue line is the simulation result, and the orange one the theoretical estimate using Eq.~(\ref{FP1}). (b): signal $X_t$ versus time $t$. (c): spectrum of the signal obtained using the fast Fourier transform, where one sees a peak at the main oscillation frequency. In (d),(e),(f) we show the same quantities as in (a),(b),(c) for  the same parameters except for $K=0.2$ and $\lambda=0.6$, for which $\lambda$ is much larger than $\lambda_c$. 
        }\label{Fig1DOsc}
\end{figure}

Last, when $d=1$ and $\sigma=0$ and $\mu=0$ the above results can be rendered particularly explicit, in this case one finds 

\begin{equation}
A^2 = \varepsilon \frac{2D}{\omega_c}=\varepsilon  \frac{2D}{K},
\end{equation}
where we have used the relationship Eq. (\ref{wcK}). In this case the shift in the frequency is given by
\begin{equation}
\Delta \omega= 2\varepsilon(3-2\sqrt{2})\omega_c,
\end{equation}
which means that the frequency increases as $\lambda$ increases, as one would intuitively expect.

\section{Numerical analysis}
\label{numerics}

\subsection{One dimensional case}
\label{1DSimu}
Now, we describe a numerical analysis of the model to check our results. We use two numerical methods. The first method, which we apply for $d=1$, consists in integrating numerically the equation for the gradient of the field $u(x,t) = \partial_x\phi(x,t)$, which satisfies
\begin{equation}
\partial_t u(x,t)   = D \partial_x^2u( x,t)-\mu \ u(x,t)+ \partial_x\delta_\sigma({ x}-{X}_t).\label{ueq} 
\end{equation}
The trajectories $X_t$ can be obtained by solving this partial differential equation with time-dependent source, by in parallel evolving $X_t$ with the equation of motion at each time step. 

The numerical simulation is carried out on a region of finite size and one must be careful in solving the diffusion equation when $\mu$ is small and so the steady state solution  for $\phi(x,t)$ can be large far from the origin. However, if the particle is confined near the origin, then the steady state solution for $\phi$ far from the origin must be asymptotically the same  as that for a point-like source term $\delta_\sigma({ x-X_t})\simeq \delta(x)$.  
The steady state solution $u_s(x)$  to the resulting equation 
\begin{equation}
 D\partial_x^2 u_s -\mu\  u_s(x,t) + \delta'(x)=0
\end{equation}
is then 
\begin{equation}
u_s(x)=-\frac{1}{2D}{\rm sgn}(x)e^{-\sqrt{\frac{\mu}{D}}|x|}.
\end{equation}
This suggests that  one can impose Dirichlet conditions for $u$ at the boundaries
\begin{align}
u(x=\pm L,t)= u_s(\pm L), 
\end{align} 
which will cover the case of finite $\mu$ for which $u_s(L)$ vanishes if $L$ is large enough ($L\gg \sqrt{D/\mu}$), and the case of vanishing $\mu$, for which $u_s$ becomes constant far from the particles. In this way, one avoids the divergence of the field $\phi$ which appears in the case $\mu=0$.  

\begin{figure}
    \centering
    \includegraphics[width=7cm]{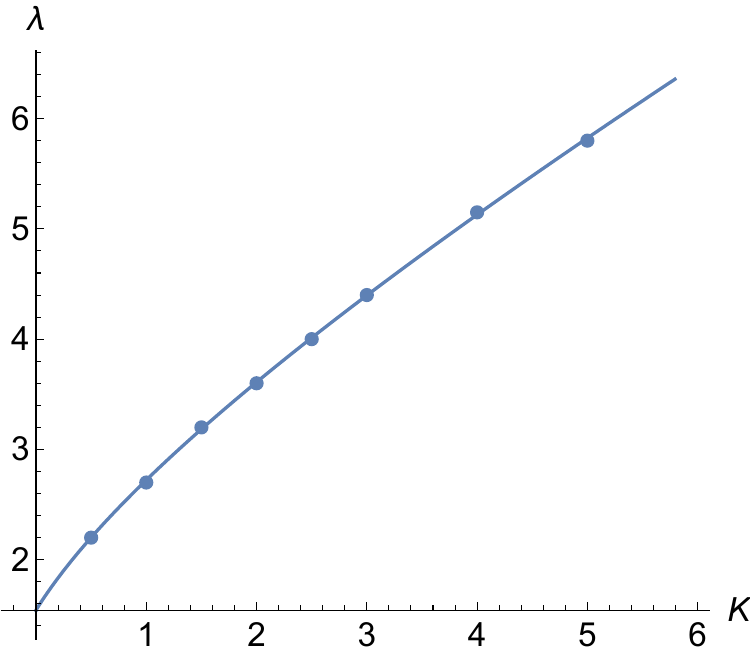}
        \caption{Line of transitions between the oscillating and non-oscillating phases for $d=1$, $D=0.5$ and $\sigma=0.2$ and $\mu=0.5$. The blue line corresponds the theoretical prediction from the stability analysis as given by Eqs.~(\ref{B1}) and (\ref{B2}). The points  correspond to the numerically determined phase boundary by examining when the onset of oscillations occurs with the method described in section \ref{1DSimu}, for $L=5$. }\label{cft}
\end{figure}

In practice, we start the simulation with this  steady state solution as the initial condition, {\em i.e.} $u(x,0)= u_s(x)$, and we also start the particle close to the origin. For $\lambda<\lambda_c$, the particle oscillates in a damped way and eventually settles down to the immobile steady state solution $X=0$ . As $\lambda$ is increased beyond $\lambda_c$, but staying close to $\lambda_c$, the particle starts to oscillate with small amplitude, see Fig.~\ref{Fig1DOsc}(b), and then settles into a steady state with an almost perfect form 
\begin{equation}
X_t= 2A \cos(\omega t +\phi),\label{eqA},
\end{equation}
as can be seen at late times in  Fig.~\ref{Fig1DOsc}(b) and also from the elliptical trajectories in the $(X_t,\dot{X}_t)$ plane on Fig.~\ref{Fig1DOsc}(a). In addition, close to the transition, there are  only two  peaks in the Fourier spectrum of $X_t$ (corresponding to $\pm\omega$) as shown in  Fig.~\ref{Fig1DOsc}(c).

 The transition is found be continuous for the amplitude $A$ which starts at zero at the phase boundary between the active and inactive phase and increases as $\lambda$ is increased. The slow temporal increase in the amplitude seen in Fig.~\ref{Fig1DOsc}(b) can be understood as being due to an exponentially growing mode due to the pole below the real axis which is however close to the real axis.

In Fig.~\ref{Fig1DOsc}(e), we show simulations when $\lambda$ is much larger than $\lambda_c$. Clearly, the amplitude   $A$ is larger, the steady state solution $X(t)$ remains periodic but  no longer with a single frequency component of the form Eq. (\ref{eqA}), as seen from the presence of several peaks in the Fourier spectrum on Fig.~\ref{Fig1DOsc}(f) and the non-elliptical shape of the trajectories in phase space, see Fig.~\ref{Fig1DOsc}(d). Interestingly, the shape of these oscillations is similar to that of stick-slip oscillations \cite{batista1998bifurcations,carlson1996}, and oscillations of molecular motor assemblies \cite{Juelicher1997PRL,Placais2009,Guerin2011b}. 

In Fig.~\ref{cft}, we show the numerically obtained phase boundary for the case $D=0.5$ and $\sigma=0.2$ and $\mu=0.5$ compared with the predictions of the stability analysis performed here. We see that the agreement is perfect (the accord is of the order of $1\%$ given potential numerical imprecisions and the fact that the decay times of the initial perturbation become very long close to the transition).

\begin{figure}
    \includegraphics[width=\linewidth]{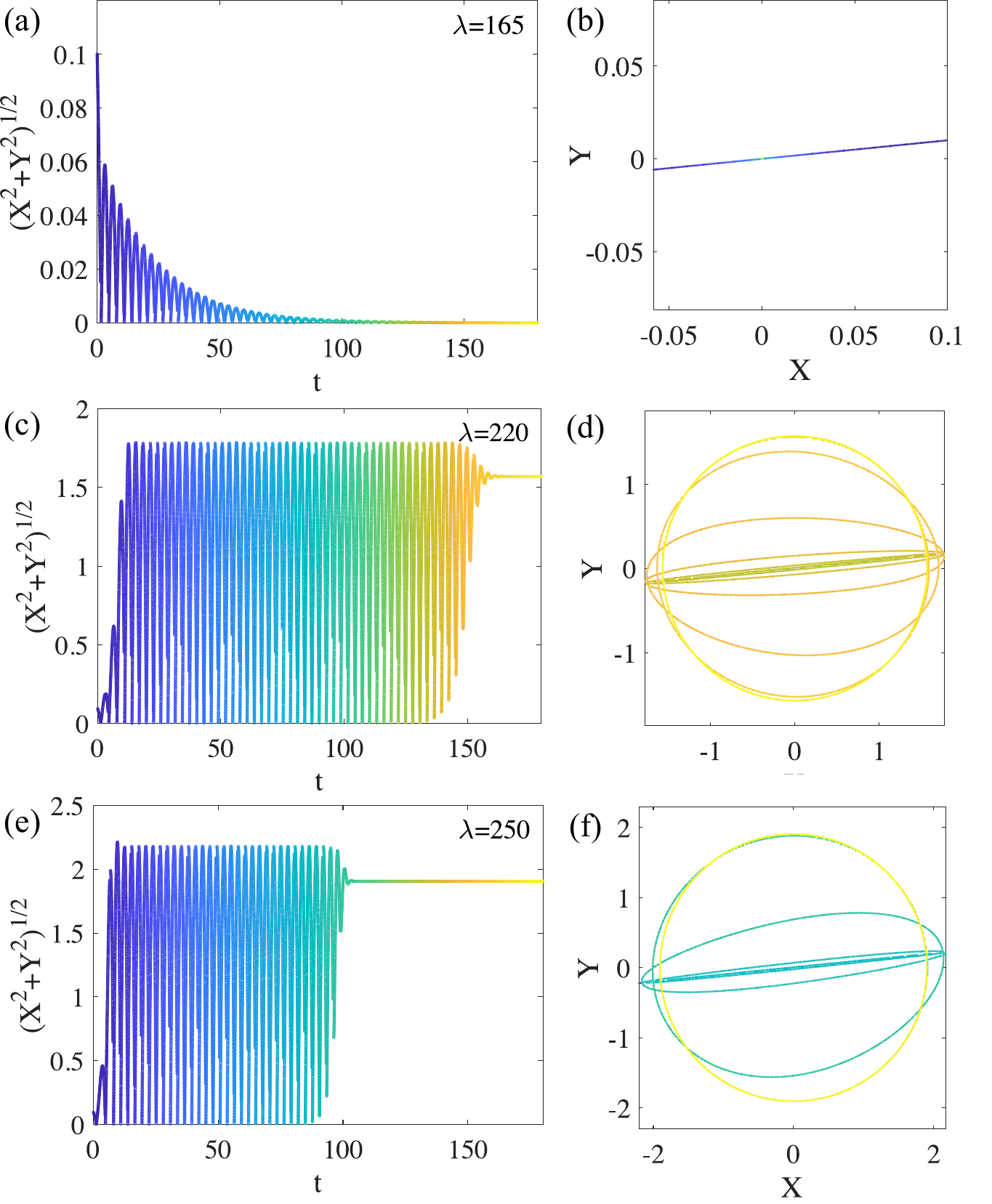}
    \caption{Oscillations for $d=2$ obtained numerically with the method of described in section \ref{SecNum2D}, for $D=\sigma=\mu=1$, $K=0.521$. For these parameters, the theoretical threshold is   $\lambda_c\simeq170.0$. In (a) 
we show trajectories as a function of time for $\lambda=165<\lambda_c$, and  the evolution of the system in the $(X,Y)$ plane. In (c) and (d), the same quantities are represented for  $\lambda=220>\lambda_c$. In (e) and (f), the same quantities are represented further away from the threshold, for  $\lambda=250$. 
Here, the time step is $d t=0.003$, $D_p=0$, and colors code (arbitrarily) for the time.  
    }  \label{FigExOscillations}
\end{figure}

\subsection{Two-dimensional case}
\label{SecNum2D}
To investigate the case $d=2$, for which $\ve[X]_t=(X_t,Y_t)$, we integrate numerically the equation of motion (\ref{EqMotion}), where the integral over the past trajectory is evaluated with Euler's method for all $\tau<t$. We do not consider any contribution from $\tau>t$ in this integral, which corresponds to the case that the activity coefficient $\lambda$ vanishes for negative times.  The initial position is $\ve[X]_0$, and lies in the vicinity of the origin. In this way, initially the motion takes place on the line joining $\ve[0]$ and $\ve[X]_0$. When $\lambda<\lambda_c$, Figs.~\ref{FigExOscillations}(a),(b) show a relaxation of $\ve[X]_t$ towards the origin, as expected. For $\lambda>\lambda_c$, we observe first a time lag where oscillations occur along a single line in the $(X,Y)$ plane, followed by a phase with circular oscillations, see Figs.~\ref{FigExOscillations}(c) to (f). This corresponds to the behavior visualized in     Fig.~\ref{TrajEnvelope}, with the system flowing close to the fixed point of straight oscillations, and then moving to the final stable fixed point of circular oscillations. As expected, the time period during which one observes straight oscillations is longer when one approaches the threshold. 

\begin{figure}
    \includegraphics[width=\linewidth]{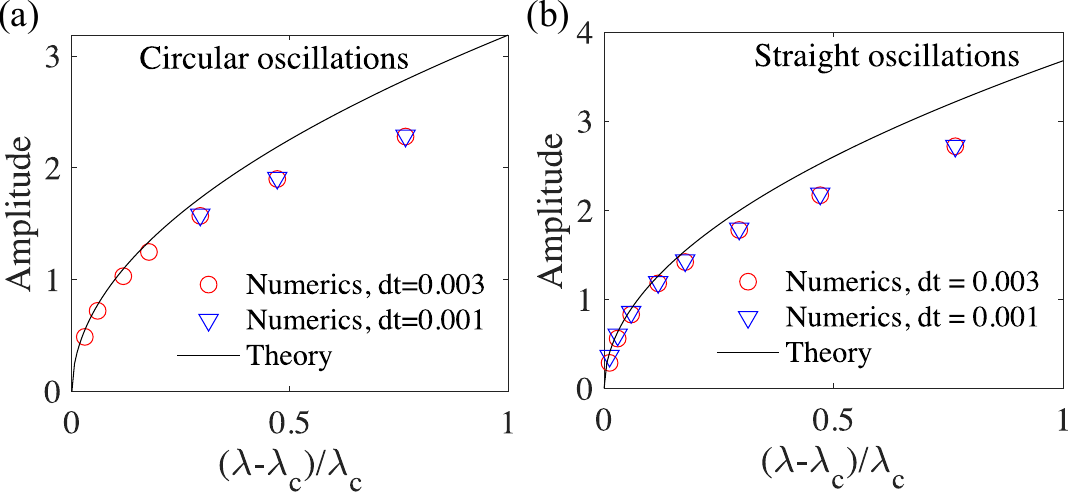}
    \caption{Amplitude of two-dimensional oscillations in the regime of circular oscillations (a) and the (temporary) regime of straight oscillations (b), defined as the maximal value of $(X_t^2+Y_t^2)^{1/2}$ over time. Symbols: simulations, the time step $dt$ is indicated in the legend. Lines are theoretical predictions using  Eqs.~(\ref{FP1}) and  (\ref{FP2}), noting that the amplitude defined in this way is $2\sqrt{R}$ for straight oscillations (with a phase shift of $n\pi$ between $X_t$ and $Y_t$) and $\sqrt{2R}$ for circular ones (with a phase shift of $\pi/2+n\pi$ between $X_t$ and $Y_t$). Parameters: $D=\sigma=\mu=1$, $K=0.521$, as in Fig.~  \ref{FigExOscillations}. 
    }  \label{FigAmplitude}
\end{figure}

To validate further our theory, we measure the amplitude of the oscillations as a function of $\lambda$, both for circular oscillations and straight ones. The results   in Figs.~\ref{FigAmplitude}(a) and (b) show a good agreement with the theory when $(\lambda-\lambda_c)/\lambda_c\ll1$, validating our approach.

\section{Conclusion}
\label{SecConclusion}

We have analyzed a simple model of an overdamped particle in a harmonic trap subject to a self phoretic driving mechanism. In agreement with results for free particles and particles confined in finite domains in similar systems \cite{lie19,koy16,koy15,koy19,miy17,nag04,nis15}, the particle undergoes a continuous transition from a stationary to an oscillating phase. This means that one can identify the phase boundary between the oscillating and immobile phase, as well as the amplitude and frequency of the oscillations close to the transition. We  determined the shape of oscillations for the two-dimensional case, finding that circular oscillations are the only stable state, but that the system can also display temporary state with straight-line oscillations. Similar behavior was observed experimentally and predicted theoretically with a simplified analysis for camphor boats in a bounded disk~\cite{koy19}. 
We have also determined fully explicit formulas describing the oscillation threshold and frequency  in the case of one-dimensional point-like particles. Our analysis is exact and does not rely on the use of simplified models of active particles or on the use a fast-varying assumption for $\phi$. It would be interesting to extend this study to bounded systems of the kind studied in  \cite{koy16,koy19}, where inertia is also taken into account, and confinement comes from the finiteness of the domain (with reflecting conditions for $\phi$ at the boundaries) instead of coming from an harmonic trap. In this case,  time derivative expansion only yields approximative results for the phase boundary but our approach may lead to exact results, as soon as one replaces the   propagator of the diffusion operator in free space by the propagator in confined domains in Eq.~(\ref{EqMotion}). It would also be possible to characterize the shape of oscillations, for which non-linearities in $\ve[X]_t$ and $\ve[X]_{t-\tau}$ are not the same as those obtained here. Another extension would be to consider non axisymmetric domains. Other possible extensions are to the case of two, or more body, systems although we note that it is not certain that such systems exhibit a continuous transition.

\begin{acknowledgements}
 T. G. acknowledges the support of the grant  ANR-21-CE30-0020. D.S. D. acknowledges  support of the grant  ANR-23 EDIPS and the Kavli Institute of Theoretical Sciences, Beijing, where part of this work was carried out. We thank Pierre Flurin and Matthieu Mangeat for early investigations on a similar problem without confinement. 
\end{acknowledgements}

\appendix

\section{Dynamic system  for $U_t$ and $P_t$}
\label{EqUAndP}
Here we give some details about how to obtain and analyse the dynamics system for $U_t$ and $R_t$. First, we define
\begin{align}
&P_t=\ve[A]_t\cdot\ve[A]_t& R_t=\ve[A]_t\cdot\ve[A]_t^*.
\end{align}
Multiplying Eq.~(\ref{EqAt}) by $\ve[A]_t$ leads to
\begin{align}
&\frac{1}{2}\dot{P}_t = \frac{\lambda_c}{1- \lambda_c \mathcal{Q}_c}\left[ \varepsilon \mathcal{G}_c P_t   - 3   \mathcal{W}_c  P_t R_t  \right].
\end{align}
Next, multiplying (\ref{EqAt}) by $\ve[A]_t^*$ and adding  the complex conjugate of the resulting expression  leads to 
\begin{align}
\dot{R}_t =  &\frac{\lambda_c}{1-\lambda_c \mathcal{Q}_c}\left[ \varepsilon   \mathcal{G}_c R_t -   \mathcal{W}_c  (P_tP_t^*   +2R_t^2)  \right] +c.c. 
\end{align}
Let us now define 
\begin{align}
M_t=P_t P_t^*. 
\end{align}
Then
\begin{align}
&\frac{1}{2}\dot{M}_t = \frac{\lambda_c}{1-\lambda_c  \mathcal{Q}_c}\left[ \varepsilon  \mathcal{G}_c M_t  - 3    \mathcal{W}_c  M_t R_t  \right] +c.c.
\end{align}
Using the real coefficients $\alpha$ and $\beta$ defined in Eq.~(\ref{DefAlphabeta}), we obtain
\begin{align}
&\frac{1}{2}\dot{M}_t =M_t[\varepsilon\alpha     - 3 \beta    R_t  ],\\
&\dot{R}_t =  \varepsilon \alpha  R_t - \beta  [M_t   +2R_t^2].   
\end{align}
Finally, using $U_t=\sqrt{M_t}$, we obtain
\begin{align}
&\dot{U}_t=\frac{\dot{M}_t}{2\sqrt{M}}=U_t[\varepsilon\alpha     - 3 \beta    R_t  ],\\
&\dot{R}_t =  \varepsilon \alpha  R_t - \beta  [U_t^2   +2R_t^2].    
\end{align}

The dynamic system for $(R_t,U_t)$ has three fixed points. The first fixed point is $(R_t=0;M_t=0)$ and corresponds to a non-oscillating state. The second fixed point is $U=R;R=\varepsilon \alpha/(3\beta)$. To determine the shape of the trajectories for this second fixed point, let us consider the two-dimensional case ($d=2$), with steady state
\begin{align}
\ve[A]_t=(\rho_x e^{i \Delta \omega t} ; \rho_y e^{i \Delta \omega t+i\varphi} ) 
\end{align}
with $\rho_x>0$ and $\rho_y>0$, and $\varphi$ a constant value at steady state. In that case, $R_t=\rho_x^2+\rho_y^2$ is fixed, and the equality $U_t=R_t$ becomes 
\begin{align}
(\rho_x^2 +\rho_y^2 e^{2i\varphi})(\rho_x^2 +\rho_y^2 e^{-2i\varphi})=(\rho_x^2+\rho_y^2)^2,
\end{align}
so that

\begin{align}
 \rho_x^2 \rho_y^2 2\cos(2\varphi)= 2\rho_y^2\rho_x^2.
\end{align}
This implies $\cos(2\varphi)=1$ so that $\varphi=0$ or $\varphi=\pi$, in both cases the trajectories in the $(x,y)$ plane take place along a segment in the two-dimensional plane. 

Last, the third fixed point is $U=0$ and $R=\varepsilon \alpha/(2\beta)$. For this fixed point, we have
\begin{align}
 \rho_x^2 + \rho_y^2 e^{2i\varphi} = 0. 
\end{align}
Considering the imaginary part of this equation leads to $\sin(2\varphi)=0$, meaning that $\varphi=n\pi/2$ for some integer $n$. Clearly, $n=0$ or $n=2$ is not acceptable (otherwise $\rho_x^2+\rho_y^2=0$). With $n=1$ or $n=3$ we have 
\begin{align}
\rho_x^2=\rho_y^2.
\end{align}
So, this third fixed point corresponds to \textit{circular oscillations} in the two-dimensional $(x,y)$ plane.

\end{document}